\documentclass[aps,showpacs,10pt]{revtex4}
\usepackage{amsmath,amssymb,bm,hyperref}
\usepackage{color}
\newenvironment{hilite}{}{}
\newcommand\Hilite[1]{#1}
\begin{document}
\title{%
  On the Evolution of Sum Rules for\\
  T-Odd Distribution and Fragmentation Functions
}


\author{\firstname{Philip G.} \surname{Ratcliffe}}
\email{philip.ratcliffe@uninsubria.it}
\affiliation{%
  Dipartimento di Scienza e Alta Tecnologia,\\
  Universit\`{a} degli Studi dell'Insubria,\\
  via Valleggio 11, 22100 Como, Italy\\
  and\\
  Istituto Nazionale di Fisica Nucleare---sezione di Milano--Bicocca,\\
  piazza della Scienza 3, 20126 Milano, Italy
}

\author{\firstname{Oleg V.} \surname{Teryaev}}
\email{teryaev@theor.jinr.ru}
\affiliation{%
  Bogoliubov Laboratory of Theoretical Physics,\\
  Joint Institute for Nuclear Research,\\
  Dubna 141980, Russia
}

\date{\today}

\begin {abstract}
  We test stability against probabilistic evolution of sum rules for transverse-momentum-dependent distribution and fragmentation functions. We find that preservation of the Burkardt sum rule for Sivers distribution functions is similar to the conservation of longitudinal momentum related to spin-averaged parton distributions. At the same time, preservation of the Schaefer--Teryaev sum rule for Collins functions is similar to preservation of the Burkhardt--Cottingham sum rule for the spin-dependent $g_2$ structure function.
\end{abstract}

\maketitle
%——————————————————————————————————————————————————————————————————————————————%
\section{Introduction}

Sum rules provide important model-independent constraints on the non-perturbative ingredients of QCD factorisation. One might mention here the charge-conservation sum rules for the spin-averaged valence-parton distributions, the momentum sum rule for the singlet case and the Bjorken sum rule for the spin-dependent distributions.

Recently, much attention has been dedicated to T-odd transverse-momentum-dependent (TMD) distribution and fragmentation functions, the best known being the Collins fragmentation and the Sivers distribution functions. They are also constrained by the sum rules that arise owing to fact that the average transverse momenta either of hadrons in a parton or of partons in a hadron are precisely zero \cite{Schafer:1999kn, Burkardt:2004ur} (note also the field-theoretical derivation \cite{Meissner:2010cc} of such sum rules for fragmentation functions).

A crucial issue is the compatibility of such sum rules with respect to QCD evolution and their resulting stability. For the evolution \cite{DGLAP} of transverse-momentum-integrated distributions, this property can be proved within the framework of the operator product expansion, whereby operators corresponding to conserved quantities are not renormalised.
However, as far as TMD evolution \cite{Aybat:2011zv} is concerned, it is still an open question. In the present paper we explore this issue using the approach to QCD evolution based on branching processes \cite{Ceccopieri:2005zz}, which is currently not proved rigorously but matches \cite{Ceccopieri:2014qha} the standard approach \cite{Aybat:2011zv} in some limit~\cite{Kodaira:1981nh}.

%-------------------------------------------------------------------------------
Let us also note that the Sivers function does not have a definite twist \cite{Ratcliffe:2007ye} while its moment, entering the Burkardt sum rule, is of twist~3. Twist~3, in turn, was the first example \cite{Efremov:1984ip} of the generation of a single-spin asymmetry by initial- or final-state interactions, described by quark--gluon correlators. The important particular case of such a correlator, the so-called gluonic pole \cite{Qiu:1991pp}, was found to be related to the relevant moment of the Sivers function, at the level of both matrix elements \cite{Boer:2003cm} and observable cross-sections \cite{Ji:2006ub} (\emph{cf}.\ also Ref.~\cite{Teryaev:2005bp}).

%-------------------------------------------------------------------------------
\section{%
  Branching, evolution of transverse moments of Sivers functions and the Burkardt sum rule
}

Let us first consider the implementation of QCD evolution for TMD functions due to partonic branching processes \cite{Ceccopieri:2005zz} in the case of T-odd distribution and fragmentation functions. More precisely, we shall address the specific form of the transverse-momentum dependence contained in these functions, which starts with the first power of transverse momentum
\begin{equation}
  \mathcal{F}_T(x_B,Q^2,\bm{k}_\perp) = k^l_\perp \, f_T^l(x_B,Q^2,k^2_\perp).
  \label{def0}
\end{equation}
A notable example is provided by the Sivers distribution function, which may be expressed in terms of the (purely transverse) two-dimensional Levi--Civita tensor $\varepsilon_\perp^{lm}$ thus
\begin{equation}
  \mathcal{F}_S^i(x_B,Q^2,\bm{k}_\perp,\bm{S}_\perp) =
  \frac{\varepsilon_\perp^{lm} k^l_\perp S^m_\perp}{M} \,
  f_S^i(x_B,Q^2,k^2_\perp),
  \label{defs}
\end{equation}
where $\bm{S}_\perp$ is the hadron transverse polarisation.
This function coincides with $f^\perp_{1T}(x_B,Q^2,\bm{k}_\perp)=f_S(x_B,Q^2,k^2_\perp)$.

We start with the generic evolution equation~\cite{Ceccopieri:2005zz}:
\begin{equation}
  \frac{\partial}{\partial \ln Q^2} \, \mathcal{F}_P^i(x_B,Q^2,\bm{k}_\perp) =
  \frac{\alpha_s(Q^2)}{2\pi}
  \int_{x_B}^1 \frac{du}{u^3} \, P_{ji}\big(u,\alpha_s(Q^2)\big)
  \int \frac{d^2\bm{q}_\perp}{\pi} \, \delta\big((1-u)Q^2-q^2_\perp\big) \,
  \mathcal{F}_P^j
    \Big( \frac{x_B}{u}, Q^2, \frac{\bm{k}_\perp-\bm{q}_\perp}{u} \Big),
  \label{dglap_TMD_space}
\end{equation}
where $i$ and $j$ label the various parton types.
Its applicability to the Sivers function is supported, first of all, by the fact that the latter appears in the tensor decomposition of the standard vector--quark correlator. This results in the presence of the unpolarised twist-2 kernel in the evolution equations under study here.
However, rigorous analysis requires account to be taken of the effective nature of the Sivers function, which also results in contributions from higher-twist operators. This leads to important modifications: in particular, those due to colour factors~\cite{Ratcliffe:2007ye}.

Substituting (\ref{defs}) into both sides of (\ref{dglap_TMD_space}) leads to
\begin{multline}
  \frac{\varepsilon_\perp^{lm} k^l_\perp S^m_\perp}{M}
  \frac{\partial}{\partial \ln Q^2} \, f_S^i(x_B,Q^2,k^2_\perp) =
  \frac{\alpha_s(Q^2)}{2\pi}\int_{x_B}^1 \frac{du}{u^3} \,
  P_{ji}\big(u,\alpha_s(Q^2)\big)
\\
  \times
  \int \frac{d^2\bm{q}_\perp}{\pi} \, \delta\big((1-u)Q^2-q^2_\perp\big) \,
  \frac{\varepsilon_\perp^{lm} (k-q)^l_\perp S^m_\perp}{u M} \,
  f_S^j
    \Big( \frac{x_B}{u}, Q^2, \frac{(\bm{k}_\perp-\bm{q}_\perp)^2}{u^2} \Big).
  \label{Sivev}
\end{multline}
Rotational invariance guarantees that the second term of $(k-q)^l_\perp$ on the r.h.s.\ is proportional to the first after integration. Projecting out with the more convenient representation
\begin{equation*}
  \bm{k}_\perp \!\cdot \bm{q}_\perp =
  {\tfrac12}\left[ k^2_\perp + q^2_\perp - (\bm{k}_\perp-\bm{q}_\perp)^2 \right]
\end{equation*}
leads to
\begin{multline}
  \int \frac{d^2\bm{q}_\perp}{\pi} \, \delta\big((1-u)Q^2-q^2_\perp\big) \,
  \frac{\varepsilon_\perp^{lm} q^l_\perp S^m_\perp}{u M} \,
  f_S^j
    \Big( \frac{x_B}{u}, Q^2, \frac{(\bm{k}_\perp-\bm{q}_\perp)^2}{u^2} \Big) =
\\
  \int \frac{d^2\bm{q}_\perp}{\pi} \, \delta\big((1-u)Q^2-q^2_\perp\big) \,
  \frac{
    \bm{k}_\perp \!\cdot \bm{q}_\perp \,
    \varepsilon_\perp^{lm} k^l_\perp \, S^m_\perp
  }{k^2_\perp u M} \,
  f_S^j \Big( \frac{x_B}{u}, Q^2, \frac{(\bm{k}_\perp-\bm{q}_\perp)^2}{u^2} \Big)
  \label{Pr}
\end{multline}
and, on substituting (\ref{Pr}) into (\ref{Sivev}), one obtains the following TMD evolution equation:
\begin{multline}
  \frac{\partial}{\partial \ln Q^2} \, f_S^i(x_B,Q^2,k^2_\perp) =
  \frac{\alpha_s(Q^2)}{2\pi} \int_{x_B}^1 \frac{du}{u^4} \,
  P_{ji}\big(u,\alpha_s(Q^2)\big)
\\
  \times
  \int \frac{d^2\bm{q}_\perp}{\pi} \, \delta\big((1-u)Q^2-q^2_\perp\big) \,
  \frac{k^2_\perp - q^2_\perp + (\bm{k}_\perp-\bm{q}_\perp)^2}{2 k^2_\perp} \,
  f_S^j
    \Big( \frac{x_B}{u}, Q^2, \frac{(\bm{k}_\perp-\bm{q}_\perp)^2}{u^2} \Big).
\end{multline}
To avoid the singularity on the r.h.s.\ as $\bm{k}_\perp\to0$, one may introduce $g_S^i(x_B,Q^2,k^2_\perp)\equiv k^2_\perp f_S^i(x_B,Q^2,k^2_\perp)$, which enters the Burkardt sum rule and which will be the main subject of our investigation. It thus satisfies the equation
\begin{multline}
  \frac{\partial}{\partial \ln Q^2} \, g_S^i(x_B,Q^2,k^2_\perp) =
  \frac{\alpha_s(Q^2)}{2\pi}\int_{x_B}^1 \frac{du}{u^2} \,
  P_{ji}\big(u,\alpha_s(Q^2)\big)
\\
  \times
  \int \frac{d^2\bm{q}_\perp}{2\pi} \, \delta\big((1-u)Q^2-q^2_\perp\big) \,
  \left[ 1 + \frac{k^2_\perp - q^2_\perp}{(\bm{k}_\perp-\bm{q}_\perp)^2} \right]
  g_S^j
    \Big( \frac{x_B}{u}, Q^2, \frac{(\bm{k}_\perp-\bm{q}_\perp)^2}{u^2} \Big).
\end{multline}

Let us now consider evolution of the second moment of the Sivers function. This enters the sum rule and is related \cite{Boer:2003cm} to twist-3 gluonic poles.
For this purpose, we first multiply both sides by $k^2_\perp$ and then integrate over the entire transverse-momentum plane. On introducing the natural variable $\bm{p}_\perp\equiv\bm{k}_\perp-\bm{q}_\perp$ into the r.h.s., so that
\begin{equation*}
  k^2_\perp - q^2_\perp + (\bm{k}_\perp-\bm{q}_\perp)^2 =
  2(p^2_\perp + \bm{p}_\perp \!\cdot \bm{q}_\perp),
\end{equation*}
one arrives at the following equation:
\begin{multline}
  \int d^2\bm{k}_\perp k^2_\perp \,
  \frac{\partial}{\partial \ln Q^2} \, f_S^i(x_B,Q^2,k^2_\perp) =
  \frac{\alpha_s(Q^2)}{2\pi} \int_{x_B}^1 \frac{du}{u^4} \,
  P_{ji}\big(u,\alpha_s(Q^2)\big)
\\
  \times
  \int \frac{d^2\bm{q}_\perp}{\pi} \, \delta\big((1-u)Q^2-q^2_\perp\big)
  \int d^2\bm{p}_\perp \, \big(p^2_\perp + \bm{p}_\perp\!\cdot\bm{q}_\perp\big)
  f_S^j \Big( \frac{x_B}{u}, Q^2, \frac{p^2_\perp}{u^2} \Big).
  \label{Sivevm}
\end{multline}
Here, in contrast with (\ref{Sivev} \& \ref{Pr}), the term linear in $\bm{q}_\perp$ integrates to zero and, after a natural rescaling $\bm{p}_\perp\to\bm{p}_\perp/u$, the evolution equation for the moment takes on the form
\begin{equation}
  \frac{\partial}{\partial \ln Q^2}
  \int d^2\bm{k}_\perp k^2_\perp f_S^i(x_B,Q^2,k^2_\perp) =
  \frac{\alpha_s(Q^2)}{2\pi}\int_{x_B}^1 \frac{du}{u}
  \Big[ u P_{ji}\big(u,\alpha_s(Q^2)\big) \Big]
  \int d^2\bm{p}_\perp \, p^2_\perp \,
  f_S^j \Big( \frac{x_B}{u}, Q^2, p^2_\perp \Big).
  \label{Sivevmf}
\end{equation}
We see that it is the DGLAP kernel $uP(u)$, describing the \emph{momentum}-density evolution, that appears here. This is quite natural, as this moment describes the average magnitude of the transverse momentum emerging owing to its correlation with a spin.

Note that evolution \cite{Braun:2009mi} of twist-3 gluonic poles at large $x$ is also reduced to multiplicative evolution with the unpolarised kernel \cite{Ratcliffe:2007ye}, which renders the results obtained compatible with the rigorous twist-3 analysis, at least in that limit. Note also that in deriving this equation no assumption has been made as to the nature of the parton distributions entering either the l.h.s.\ or r.h.s. This means that similar equations are also valid for the contributions of gluonic TMDs to the evolution of quark distributions as well as for both the gluon and quark contributions to the evolution of gluonic TMDs. As a result, the indices $i$ and $j$ take on the values $i,j=q,G$ (for singlet quark and gluon distributions, respectively). Finally, the \emph{first} $x$-moments of the transverse moments entering the Burkardt sum rule evolve like the \emph{second} moments of spin-averaged distributions. The stability of the Burkardt sum rule has the same origin as the stability of the momentum sum rule for the spin- and transverse-momentum-averaged distributions.

The asymptotic solution is however different: for the usual spin-averaged distribution, evolution leads to the (positive) asymptotic ratio (being actually not too far from the ratio ${\sim}1$ found at low and moderate $Q^2$. For the Sivers function, the positive asymptotic ratio of the (singlet) quark and gluon distributions is compatible with their opposite signs provided by the Burkardt sum rule if and only if the quark and gluon Sivers function are separately zero in the asymptotic limit. The observed smallness of the gluon Sivers functions indicates that, for some reason, moderate $Q^2$ values are also not far from the asymptotic limit.

Let us also mention that the sum rule is preserved provided integration over all $\bm{k}_\perp$ is performed. As soon as it is natural to take the input for the TMDs in Gaussian form (which may be also motivated by making an analogy to non-local vacuum condensates \cite{Teryaev:2004df, Ratcliffe:2007ye}), evolution leads to the development of a power-like ``tail''. In the current approach account for such a tail appears to be compulsory, in order to prove stability of the sum rules.
\Hilite{\relax
And we should stress, that by taking the Gaussian form of the Sivers function as input, one should still have the function with a power tail, integrable over the weight $p_\perp^2$, which would signal either a decrease of the Sivers function faster than $p_\perp^3$ or its oscillation. The latter scenario, although possible in principle, does not occur for positive Gaussian input, as may easily be demonstrated by generalising the proof \cite{Artru:2008cp} of preservation of positivity by DGLAP evolution to the case under consideration
}

%-------------------------------------------------------------------------------
\section{%
  Evolution of the transverse moment of the Collins fragmentation function and Schaefer--Teryaev sum rule
}

The evolution of fragmentation functions is also governed by branching processes, which is quite intuitive and was, in fact, established \cite{Bassetto:1979nt} earlier than the similar approach to the evolution of parton distributions explored above.
The specific form of the transverse-momentum dependence contained in the Collins function, which also starts with the first power of transverse momentum, may be expressed as
\begin{equation}
  \mathcal{D}_T^i(z,Q^2,\bm{k}_\perp)
  =
  \frac{\varepsilon_\perp^{lm} k^l_\perp S^m_\perp}{M} \, D_T^i(z,Q^2,k^2_\perp),
\end{equation}
where now $\bm{S}_\perp$ is the quark transverse polarisation and $i$ the hadron type.

The generic evolution equation~\cite{Bassetto:1979nt, Ceccopieri:2005zz}:
\begin{equation}
  \frac{\partial}{\partial \ln Q^2} \, \mathcal{D}^i_T(z,Q^2,\bm{k}_\perp) =
  \frac{\alpha_s(Q^2)}{2\pi}
  \int_{z}^1 \frac{du}{u} \, \delta{P} \big(u,\alpha_s(Q^2)\big)
  \int \frac{d^2\bm{q}_\perp}{\pi} \, \delta\big(u(1-u)Q^2-q^2_\perp\big) \,
  \mathcal{D}^i_T
    \Big( \frac{z}{u}, Q^2, \bm{k}_\perp - \tfrac{z}{u}\bm{q}_\perp \Big),
  \label{dglap_TMD_time}
\end{equation}
where $\delta{P}$ is the transversity~\cite{Artru:1990zv} evolution kernel. Its applicability to the Collins function is supported by the fact that the latter appears in the tensor decomposition of the quark correlator containing the Dirac matrix $\sigma^{\mu\nu}$.
Use of rotational invariance, analogously to the case of the Sivers function, leads to the following TMD evolution equation:
\begin{multline}
  \frac{\partial}{\partial \ln Q^2} \, D_T^i(z,Q^2,k^2_\perp) =
  \frac{\alpha_s(Q^2)}{2\pi} \int_{z}^1 \frac{du}{u} \,
  \delta{P} \big(u,\alpha_s(Q^2)\big)
\\
  \times
  \int \frac{d^2\bm{q}_\perp}{\pi} \, \delta\big(u(1-u)Q^2-q^2_\perp\big) \,
  \frac{
    k^2_\perp - (\frac{z}{u})^2 q^2_\perp
    + (\bm{k}_\perp - \frac{z}{u} \bm{q}_\perp)^2
  }{2k^2_\perp} \,
  D_T^i
    \Big(
      \frac{z}{u}, Q^2, \big(\bm{k}_\perp-\tfrac{z}{u}\bm{q}_\perp\big)^2
    \Big).
\end{multline}
As a result, the evolution of the transverse moment of the Collins function is also multiplicative. However, the DGLAP kernel $\delta{P}$ is now that of transversity evolution, owing to the presence of the above-mentioned sigma matrix in the quark correlator:
\begin{equation}
  \frac{\partial}{\partial \ln Q^2}
  \int d^2\bm{k}_\perp k^2_\perp D_T^i(z,Q^2,k^2_\perp) =
  \frac{\alpha_s(Q^2)}{2\pi} \int_{z}^1 \frac{du}{u^2}
  \Big[ u \, \delta{P} \big(u,\alpha_s(Q^2)\big) \Big]
  \int d^2\bm{p}_\perp \, p^2_\perp \,
  D_T^i \Big( \frac{z}{u}, Q^2, p^2_\perp \Big).
  \label{Collf}
\end{equation}
Note that the second moment of $\delta{P}$ is entirely unrelated to any longitudinal-momentum evolution. However, if one sums over all finite hadron species $i$, the integral on the r.h.s.\ vanishes owing to the Schaefer--Teryaev sum rule. Consequently, its derivative on the l.h.s.\ is also zero, which justifies its stability. Such preservation of sum rules under evolution due to boundary conditions is, in fact, similar to the case \cite{Braun:2001qx} of the Burkhardt--Cottingham sum rule.

\section{Conclusions}

The probabilistic evolution of TMD distribution and fragmentation functions naturally accommodates their sum rules.
While preservation of the Burkardt sum rule for the Sivers distribution functions is very similar to that of longitudinal momentum, preservation of the Schaefer--Teryaev sum rule for Collins fragmentation functions appears to have much in common with preservation of the Burkhardt--Cottingham sum rule for the $g_2$ structure function.

The stability of sum rules against the different and more rigorous evolution equations remains to be studied. This possibility is supported by the compatibility with twist-3 evolution at large~$x$. The approach adopted in \cite{Bassetto:1979nt} is often considered as interpolating between DGLAP \cite{DGLAP} and BFKL \cite{Lipatov-etal} evolution; so that the observed stability of the sum rules may be a hint as to the interplay between TMD and unintegrated parton distributions (the non-perturbative ingredients of BFKL factorisation).

\begin{acknowledgments}
We are indebted to Federico Ceccopieri for collaboration on earlier stages of this work and for his many useful comments.
We should also like to thank Anatoli Efremov for helpful discussions and reading the manuscript.

This work was supported in part by the Russian Foundation for Basic Research (Grants 13-02-01060, 12-02-00613 and 12-02-91526) and, at earlier stages, by the Cariplo foundation.
\end{acknowledgments}

\end{document}